\documentclass[12pt,preprint]{aastex}

\newcommand{\etal}{{\em et al.\ }}

\def\ion[#1 #2]{#1\,{\sc #2}}
\def\ergs[#1]{#1 {ergs}~{cm$^{-2}$}\,{s$^{-1}$}\,{sr$^{-1}$}}
\def\dens[#1]{10$^{#1}$\hskip 1.5pt{cm$^{-3}$}}
\def\densr[#1 #2]{10$^{#1}$\hskip 1pt{--}\hskip .5pt{10$^{#2}$}\hskip 1.5pt{cm$^{-3}$}}
\def\fl[#1 #2]{{#1}$\pm${#2}}
\def\orb[#1 #2]{{$#1^{#2}$}}
\def\ls[#1 #2]{{$^{#1}${#2}}}
\def\tm[#1 #2 #3]{{$^{#1}${#2}$_{#3}$}}

\begin{document}

\title{Prominence plasma diagnostics through EUV absorption}

\author{E. Landi,\altaffilmark{1}, F. Reale\altaffilmark{2} }

\altaffiltext{1}{Department of Atmospheric, Oceanic and Space Sciences, University of Michigan, Ann Arbor, MI 48109}
\altaffiltext{2}{Dipartimento di Fisica, Universit\`a di Palermo, Piazza del Parlamento 1, 90134 Palermo, Italy}

\begin{abstract}
In this paper we introduce a new diagnostic technique that uses
prominence EUV and UV absorption to determine the prominence plasma
electron temperature and column emission measure, as well as He/H 
relative abundance; if a realistic assumption on the geometry of
the absorbing plasma can be made, this technique can also yield
the absorbing plasma electron density. This technique capitalizes 
on the absorption properties of Hydrogen and Helium at different 
wavelength ranges and temperature regimes. Several cases where this 
technique can be 
successfully applied are described. This technique works best when
prominence plasmas are hotter than 15,000~K and thus it is ideally
suited for rapidly heating erupting prominences observed during the
initial phases of coronal mass ejections. An example is made using
simulated intensities of 4 channels of the SDO/AIA instrument. This 
technique can be easily applied to existing observations from almost 
all space missions devoted to the study of the solar atmosphere, which
we list.
\end{abstract}


\section{Introduction}

Prominences are a common feature of both the active and the quiet
solar corona. They consist of large structures of plasma in the solar
inner atmosphere maintained
by a strong and complex magnetic field configuration, which is able
to keep their very low temperature plasma (i.e. 10,000~K or less
in their cores) separated from the multimillion solar corona that
surrounds them. Such magnetic field configuration may last as long
as several rotations, but it can also be de-stabilized; in the latter
cases, the prominence erupts and is ejected in the interplanetary
space forming the core of a coronal mass ejection (CME).

Since prominence plasmas are very cold, they can be observed in the
visible through \ion[H i] emission (outside the limb) or as dark
features called filaments (inside the disk); absorption features are 
also a common prominence
manifestation at shorter wavelengths. Visible observations of prominences
have been carried out since the 1800s, and a large body of literature has
been produced that studied their morphology, evolution, and their physical
and dynamical properties. Reviews of these results can be found, for
example, in Labrosse \etal (2010), Tandberg-Hanssen (1995) and references
therein.

One of the main open questions in prominence science is the role played
by these structures in the initiation and in the propagation of CMEs.
In fact, prominences are present in a sizeable fraction
of all CMEs launched by the Sun, and prominence 
plasma has been also observed in-situ by mass spectrometers carried by 
the ACE, Ulysses and STEREO satellites. In-situ measurements of prominence 
plasma properties such as element and charge state composition can provide
very important information on the original prominence element composition:
despite being largely unknown, plasma composition can provide information 
both on the origin of the prominence plasma itself, and on the heating and 
cooling processes experienced by the prominence during the early phases of 
CME acceleration. For example,
Landi \etal (2010) provided evidence that erupting prominence are heated
to temperatures in excess of 200,000~K in the earliest phases of CME
initiation; still, Lepri \& Zurbuchen (2010) and Gilbert \etal (2012) 
showed that cold prominence material ($T\simeq 40-70,000$~K, where
singly ionized C, O and Fe were present) is common in interplanetary CMEs.
Understanding the thermal history of an erupting CME may shed light 
on the unknown processes that create a CME in the first place. Coordinated
studies of in-situ measurements of plasma element and charge state 
composition and remote-sensing determinations of plasma temperature 
and density during acceleration in the same CME event can provide 
vital constraints to CME initiation models, as shown by Gruesbeck 
\etal (2011,2012), and Landi \etal (2012).

Measuring the physical properties of prominences has proven to be
difficult. In the visible range, it is necessary to address radiative
transfer of the prominence emission, so that complex models are
necessary to reproduce the observations and describe the physical
properties of prominence plasmas. The recent launch of satellite-borne
X-ray, EUV and UV instrumentation has opened a new window in prominence
science that enabled to study the little known prominence-corona
transition region and its properties on one side, and EUV and X-ray
absorption on the other. 

Prominence absorption depends on the electron temperature of the 
prominence plasma when T exceeds $\simeq 15,000$~K, and thus it can 
be used to determine the electron temperature of an erupting prominence.
Also, the properties of the absorption coefficient can be used to
determine the prominence plasma Helium abundance.
In this paper we revisit the EUV and UV absorption of prominence
plasmas and develop a new diagnostic technique that utilizes the
EUV and UV absorption of a prominence (either quiescent or eruptive)
to determine its electron temperature and He/H abundance. This 
technique can be applied to observations from both high-resolution 
spectrometers and narrow-band imagers from all past and current
space missions such as SOHO, TRACE, STEREO, Hinode and SDO,
as well as future missions such as Solar-C, Solar Probe and Solar
Orbiter. The principle of this diagnostic technique is described
in Section~\ref{technique}, and a few cases where it can be applied
are outlined in Section~\ref{applications}. Section~\ref{summary}
suggests future applications to existing data.

\section{The diagnostic technique}
\label{technique}

The diagnostic technique that we have developed relies on the
EUV absorption properties of the prominence material through
bound-free transitions. Thus, it is useful to recall a few 
basic properties of bound-free EUV absorption.

\subsection{The absorption coefficient}

The intensity $F_{obs}$ emerging from a non-emitting slab with 
thickness $S$ made of material that absorbs the incident radiation
$F_{inc}$ emitted by a source located behind the slab is given by

\begin{eqnarray}
F_{obs} & = & F_{inc}~e^{-\tau}  \quad\quad\quad\quad\quad {\mathrm{where}} \label{abs_gen} \\
\tau & = & \int_0^S{n_{abs}k_{abs}dl}   \label{tau_gen}
\end{eqnarray}

\noindent
where $k_{abs}$ is the absorption coefficient and $n_{abs}$ is
the total number density of absorbers along the radiation path. 
In realistic cases, the absorbing plasma is made of several 
different elements, each distributed in a range of ionization 
stages which depend on the physical properties of the absorbing 
plasma itself. In this case, assuming that each ion in the 
absorbing plasma interacts with the incident radiation 
independently from the others, the absorption coefficient can 
be expressed as

\begin{eqnarray}
\tau & = & \int_0^S{{\left[{\sum_X{\sum_m{n{\left({X^{m+}}\right)}k_{X,m}}}}\right]}dl}
\label{tau_prec}
\end{eqnarray}

\noindent
where $n{\left({X^{m+}}\right)}$ is the density of ions $m+$ of 
the element $X$ present along the line of sight, each characterized 
by its own absorption coefficient $k_{X,m}$. In EUV observations 
of solar prominences (i.e. between $\approx$100~\AA\ and $\approx$1000~\AA), 
absorption is dominated by H and He, which are the most abundant 
components of solar plasmas. In the temperature range typical of 
prominences, a combination of \ion[H i], \ion[He i] and \ion[He ii] 
atoms and ions will be present along the line of sight, so that 
Equation~\ref{tau_prec} can be simplified to

\begin{eqnarray}
\tau & = & \int_0^S{n_Hk_{eff}dl} \label{kappa_simple}\\
k_{eff} & = & f{\left({H~I,T}\right)}k_{H~I}+A_{He}{\left[{f{\left({He~I,T}\right)}k_{He~I}+f{\left({He~II,T}\right)}k_{He~II}}\right]} \label{kappa_eff}
\end{eqnarray}

\noindent
where the functions $f$ indicate the fractional abundances of \ion[H i],
\ion[He i] and \ion[He ii] (for example, for \ion[He ii]
$f{\left({He~II,T}\right)}=n{\left({He~II}\right)}/n{\left({He}\right)}$),
which depend on the plasma electron temperature, and 
$k_{(ion)}$ is the individual absorption cross sections of each of these 
three species. $A_{He}=n{\left({He}\right)}/n{\left({H}\right)}$ represents 
the abundance of He relative to H, while $n_H$ is the total number density
of Hydrogen (included both neutral and ionized).

Equation~\ref{kappa_eff} divides the 100-1000~\AA\ wavelength range into
four main regions of interest:

\begin{eqnarray}
{\mathrm{A:}} & 912 < \lambda < 1000 & k_{eff}= 0 \nonumber \\
{\mathrm{B:}} & 504 < \lambda < 912  & k_{eff}= f{\left({H~I,T}\right)}k_{H~I} \nonumber \label{regions} \\
{\mathrm{C:}} & 228 < \lambda < 504  & k_{eff}= f{\left({H~I,T}\right)}k_{H~I}+A_{He}f{\left({He~I,T}\right)}k_{He~I} \\
{\mathrm{D:}} & 100 < \lambda < 228  & k_{eff}= f{\left({H~I,T}\right)}k_{H~I}+A_{He}{\left[{f{\left({He~I,T}\right)}k_{He~I}+f{\left({He~II,T}\right)}k_{He~II}}\right]} \nonumber
\end{eqnarray}

\noindent
which are defined by the absorption edges of each of the three species. 
Beyond 912~\AA\ no absorption is present and the prominence is largely
optically thin. Under the assumption of ionization equilibrium, the $f$
ion fractions are a known function of the electron temperature (e.g. 
Bryans \etal 2009, Dere \etal 2009 for $T_e$ beyond 10,000~K); also, 
the absorption cross sections for each species are known from atomic 
physics (e.g. from the photoionization cross sections of Verner \& 
Ferland 1996). Thus, the absorption process is 
dependent on three main, unknown properties of the absorbing material: 
its total Hydrogen number density $n_H$, its electron temperature, and 
its Helium abundance relative to Hydrogen.

\begin{figure}
\includegraphics[width=15.0cm,height=17.0cm,angle=90]{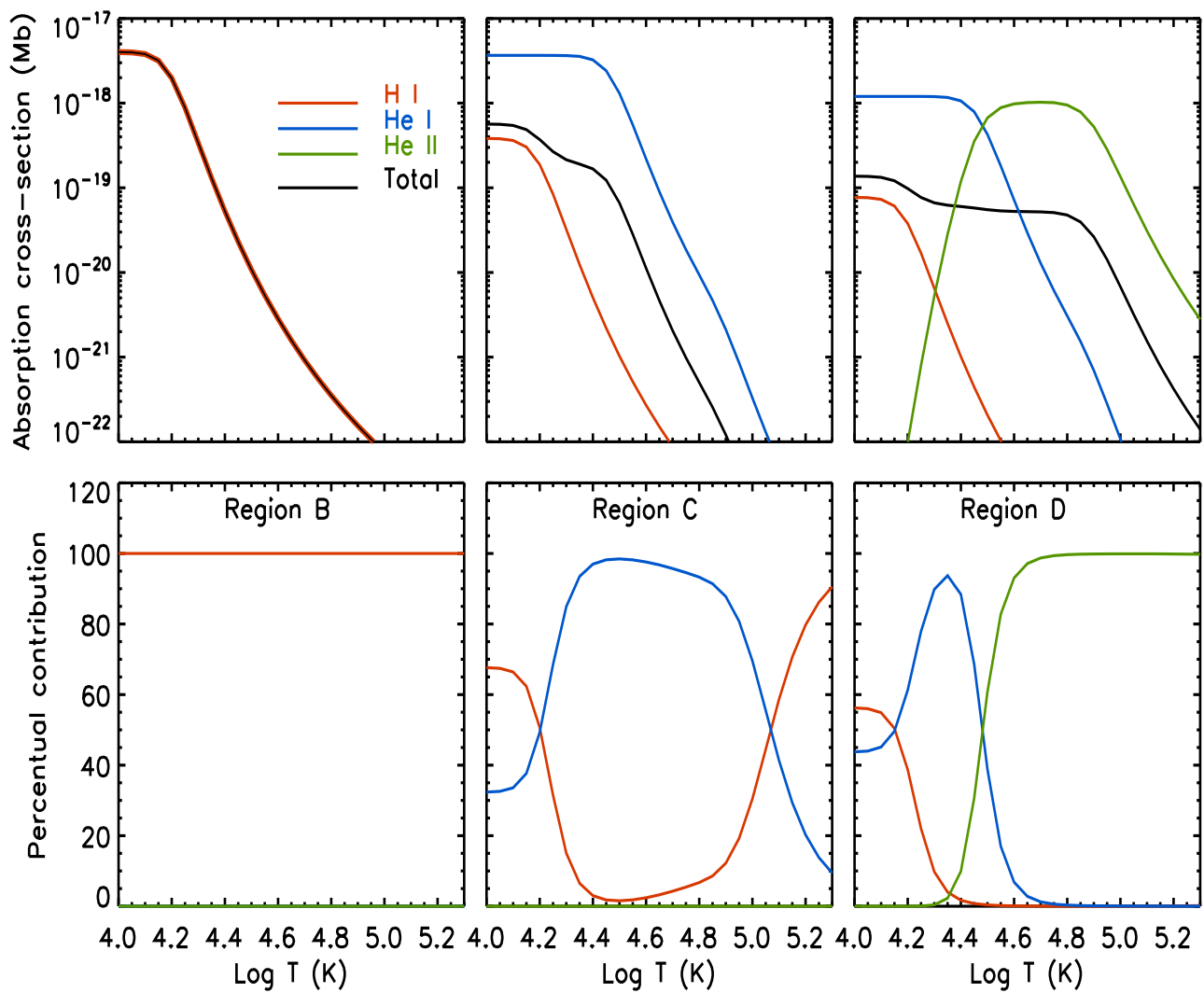}
\caption{\label{keff_5} {\bf Top:} absorption coefficient $k_{eff}$ 
as a function of temperature for wavelength regions~B, C and D (defined
in Table~\ref{combinations}). Helium abundance has been assumed to be
5\% of Hydrogen ($A_{He} = 0.05$). Red: \ion[H i]; Blue: \ion[He i]; Green: \ion[He ii];
Black: total $k_{eff}$ given by Equation~\ref{kappa_eff}.}
\end{figure}

\begin{figure}
\includegraphics[width=15.0cm,height=17.0cm,angle=90]{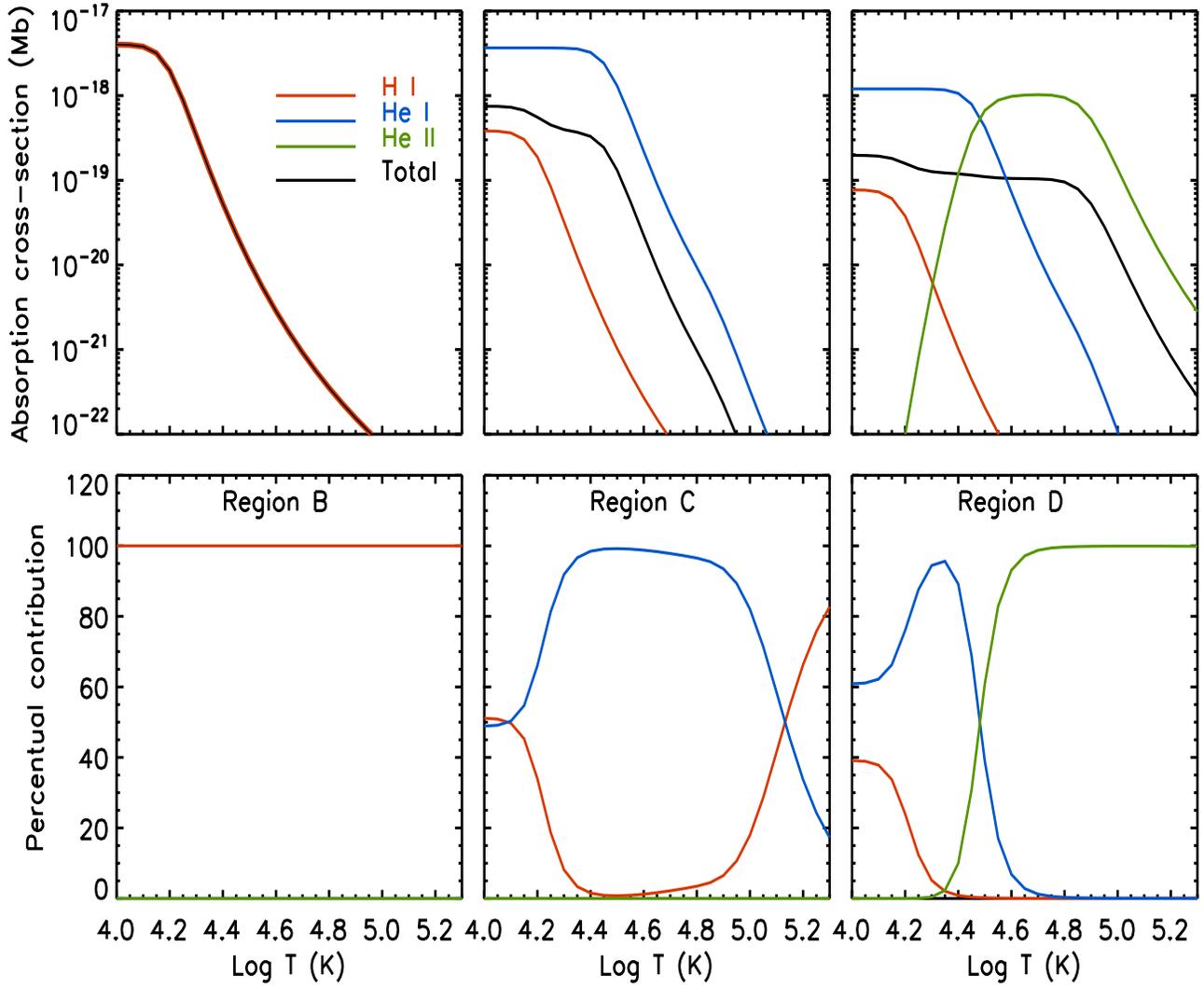}
\caption{\label{keff_10} Same as Figure~\ref{keff_5}, with $A_{He}=0.10$.}
\end{figure}

The top panels of Figures~\ref{keff_5} and \ref{keff_10} display $k_{eff}$ 
as a function of temperature for the B, C and D wavelength regimes, 
adopting the He/H values of 5\% and 10\%, respectively. The absorption
cross sections have been calculated using the photoionization cross
sections of Verner \& Ferland (1996). The bottom panels
show the percent contribution of each of the three ions to the total
value of $k_{eff}$. In Region~B absorption is only due to neutral H and 
as the plasma temperature rises and H ionizes, it decreases very quickly; 
there is no dependence on the He abundance. The temperature dependence of 
$k_{eff}$ is very strong for $T_e \geq$15,000~K, while it is very mild 
at lower temperatures where almost all Hydrogen is neutral.

In Regions~C and D the presence of He absorption has two main effects
on $k_{eff}$: first, it introduces the dependence on the He/H relative 
abundance; and second, it extends the range where $k_{eff}$ is weakly 
dependent on temperature to higher values. The latter effect is due to 
the fact that the ionization potential of He is larger than the H one, 
so that He resists ionization for a larger temperature range than H, 
and the fact that in region~D the absorption capability lost with 
\ion[He i] ionization is replaced by \ion[He ii] absorption.

Figures~\ref{keff_5} and \ref{keff_10} thus show that 1) the absorption 
properties of the slab depend on the electron temperature in different 
ways depending on the wavelength of the EUV incident radiation; and 2)
they are fairly constant up to almost 100,000~K at wavelengths below 
228~\AA. Also, above $\approx$25,000~K the slab absorption in region~D
is almost exclusively due to He rather than H (in region~B Hydrogen becomes
again important for $T>80,000$~K), whereas they are only due to H at 
wavelengths between 504~\AA\ and 912~\AA.

\subsection{EUV absorption as diagnostic tool}

The diagnostic tool we propose in the present work capitalizes on
the different temperature dependence of $k_{eff}$ in the three
different wavelength regions. In order to be applied, it requires
observations at wavelengths spanning at least two of the three
regions; availability of observations in all three regions further
increases the possible applications. Fortunately, combinations of
the available space instrumentation allows the application of these
technique to existing observations of both filaments (that is, 
prominences observed inside the solar disk) and prominences at 
the limb, as shown in Table~\ref{combinations}. Almost all 
instruments sample wavelengths at least in two of these three 
ranges.

Since the temperature effects on $k_{eff}$ are most evident at
temperature ranges larger than $\approx$15,000~K, this technique
is best applied to prominences whose plasma is relatively hot to
begin with, or is being heated, such as in erupting prominences
(e.g. Landi \etal 2010).

\subsection{Basic principle of the diagnostic technique}
 
The diagnostic technique we propose relies on the determination
of the extinction coefficient $e^{-\tau}$ for many different 
spectral lines or narrow-band filters. Let us assume for a 
moment that we have been able to measure the extinction coefficient 
$e^{-\tau}$ in Equation~\ref{abs_gen}, for example by determining 
the ratio $F_{obs}/F_{inc}$. We will discuss in the next section a 
few cases where such a measurement (or an equivalent one) can be 
done.  We will also further assume that the physical properties of 
the absorbing material are approximately the same along the path 
length $S$ of the absorbing material crossed by the line of sight. 
In this case, Equations~\ref{abs_gen} and \ref{kappa_simple} can 
be combined to give

\begin{eqnarray}
\tau & = & k_{eff}{\left({A_{He},T}\right)}N_H \label{column_density} \\
\ln {\left({F_{obs}/F_{inc}}\right)} & = & -k_{eff}{\left({A_{He},T}\right)}N_H \label{main_eq}
\end{eqnarray}

\noindent
where $N_H=\int_0^S{n_Hdl}$ is the Hydrogen column density (in
cm$^{-2}$). It is important to notice that $N_H$ and $S$ are 
properties of the absorbing plasma only.

The effective absorption coefficient $k_{eff}$ can be calculated
as a function of temperature for any spectral line or narrow band
imaging channel, once $A_{He}$ has been specified. In case we have
a narrow-band filter, $k_{eff}$ can be easily calculated as it 
changes slowly with wavelength over the width of the filter itself,
which usually encompasses from a few to a few tens of \AA\ (within
$\approx 20$\% in the available narrow-band imagers). This allows 
us to define and calculate as a function of temperature the function

\begin{eqnarray}
L{\left({T}\right)} = {{1}\over{k_{eff}{\left({T}\right)}}} \ln{\left({{{F_{obs}}\over{F_{inc}}}}\right)} = {{k_{eff}{\left({T_{abs}}\right)}}\over{k_{eff}{\left({T}\right)}}} N_H \label{lfunction}
\end{eqnarray}

\noindent
where we indicate with $T_{abs}$ the temperature of the absorbing
material. The main property of the $L{\left({T}\right)}$ function
is 

\begin{equation}
L{\left({T_{abs}}\right)} = N_H \label{main_property}
\end{equation}

\noindent
for any spectral line or narrow-band imaging channel we consider.
Since $N_H$ is a property of the absorbing material only, the
$L{\left({T_{abs}}\right)}$ values of all the spectral lines or
narrow band filters are the same.
Equation~\ref{main_property} allows us to use the function
$L{\left({T_{abs}}\right)}$ in the same way as the L-function
defined by Landi \& Landini (1997): if we measure $F_{inc}$ and
$F_{obs}$ (or some combination of them, as we will see in the next
Section) for a number of spectral lines and/or narrow-band images,
and plot their $L{\left({T_{abs}}\right)}$ functions in the same
figure as a function of temperature, all curves will cross the 
same point ${\left({T_{abs},N_H}\right)}$. The coordinates of 
the crossing point can then be used to determine both the 
absorbing plasma electron temperature and the $N_H$ value; the
latter can in turn be used to determine the average number density
of H in the absorbing material using some assumption or estimate 
of the length $S$.

Also, the presence of a single crossing point for all curves 
provides a check on the main assumption of this technique,
namely that the absorption properties of the prominence plasma
are approximately the same everywhere in the prominence itself.
Also, the behavior of the $L{\left({T_{abs}}\right)}$ functions
of different wavelength regimes can also provide a direct 
determination of $A_{He}$.

\subsection{Example: SDO/AIA narrow band images}

\begin{figure}
\centerline{\includegraphics[width=10.0cm,height=12.0cm,angle=90]{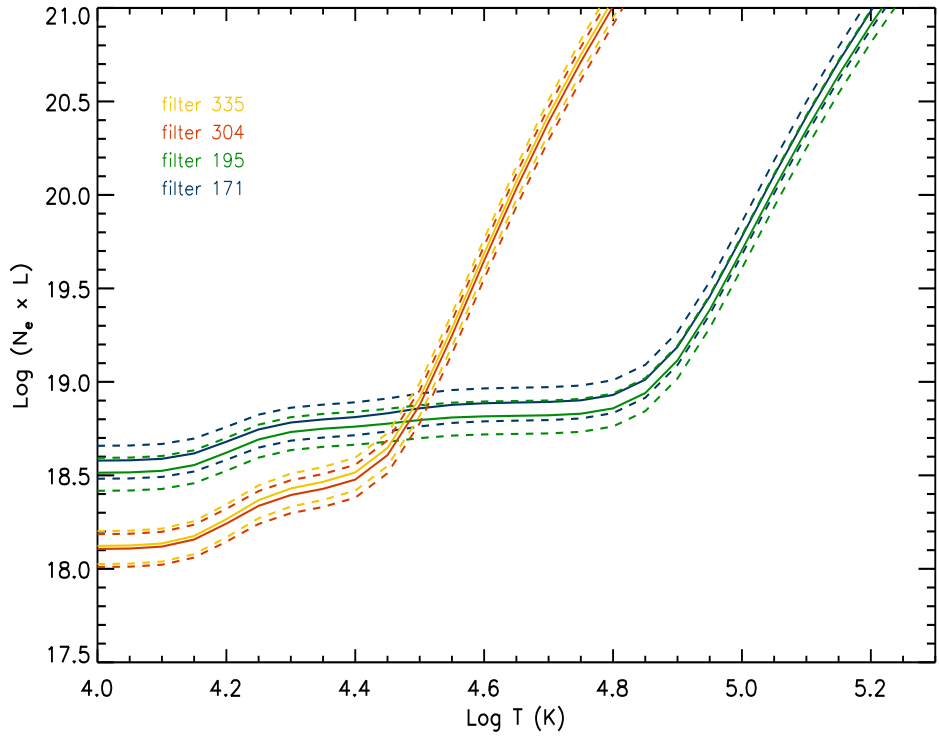}}
\caption{\label{tech_example} Example of the diagnostic technique
applied to simulated intensities for four channels of the SDO/AIA
narrow band imagers (see text for details). The crossing point allows
to determine the plasma temperature and column density.}
\end{figure}

An example of this technique is shown in Figure~\ref{tech_example}.
In this figure, simulated $F_{obs}/F_{inc}$ ratios have been calculated 
at the wavelengths of a few SDO/AIA channels for a prominence with total 
$N_H=6.3\times 10^{18}$ cm$^{-2}$, assuming constant $k_{eff}$ over the 
entire width of the filter bandbass; this corresponds to $F_{obs}/F_{inc}$ 
ratios of 0.66, 0.55, 0.56, and 0.49 for the 171~\AA, 195~\AA, 304~\AA\
and 335~\AA\ SDO channels, respectively. We added a small amount to the 
$L(T)$ functions of the 195~\AA\ and 335~\AA\ curves to make them more 
easily visible in the plot (otherwise the 304~\AA\ and 335~\AA\ curves, 
as well as the 171~\AA\ and 195~\AA\ curves, would have been coincident). 
An arbitrary 20\% uncertainty has been associated to each curve and shown
as dashed lines in Figure~\ref{tech_example}. The 304~\AA\ channel has 
also been assumed to produce zero \ion[He ii] 304~\AA\ line emission. 
The He/H abundance has been assumed to be $A_{He}=0.085$. The SDO/AIA 
channels sample wavelengths belonging to regions $C$ and $D$, so that 
their $L{\left({T}\right)}$ functions have a very different temperature 
dependence due to the \ion[He ii] absorption: as expected, their 
crossing point is very sharply defined and provides a rather 
precise measurements of $T_{abs}$ and $N_H$. On the contrary,
channels whose wavelengths belong to the same spectral region 
provide completely overlapping $L{\left({T}\right)}$ functions so 
that they can not provide a defined crossing point. However, if
their $L{\left({T}\right)}$ functions do not overlap, the difference
between them can be used to indicate one (or a combination) of the
following scenarios: 1) presence of emission from the absorbing
material itself; 2) problems in the atomic physics; 3) possible
multi-temperature structure of the absorbing prominence; 4) 
inaccuracy of the assumed $A_{He}$ value used to calculate 
$L{\left({T}\right)}$ itself.

It is important to note that when lines or channels from regions
$C$ and $D$ only are available, and the temperature of the absorbing
slab is between $\approx$15-20,000~K and $\simeq 80,000$~K, only He 
absorption is significant so that the ordinate of the crossing point 
allows the direct determination the Helium column density with no
assumption on $A_{He}$. When lines or channels from all three 
wavelength ranges are available and the temperature is in the
15,000-80,000~K range, the difference in height between region~$B$ 
$L{\left({T}\right)}$ functions and the crossing point defined by 
regions $C$ and $D$ $L{\left({T}\right)}$ functions provides a 
direct measurement of $A_{He}$. This measurement is very important 
in the case of erupting prominences as it can be compared with 
in-situ $A_{He}$ measurements from Interplanetary Coronal Mass 
Ejections (ICMEs).

\section{Applications of the technique}
\label{applications}

In order to utilize the properties of EUV absorption for prominence
plasma diagnostics, it is necessary to first define a function
$L{\left({T}\right)}$ that has the properties described in the
previous Section using observed EUV line or narrow-band intensities.
The easiest way to achieve this is to compare intensities measured 
over a prominence with their values measured near the prominence 
itself.

\begin{figure}
\includegraphics[width=15.0cm,height=7.0cm]{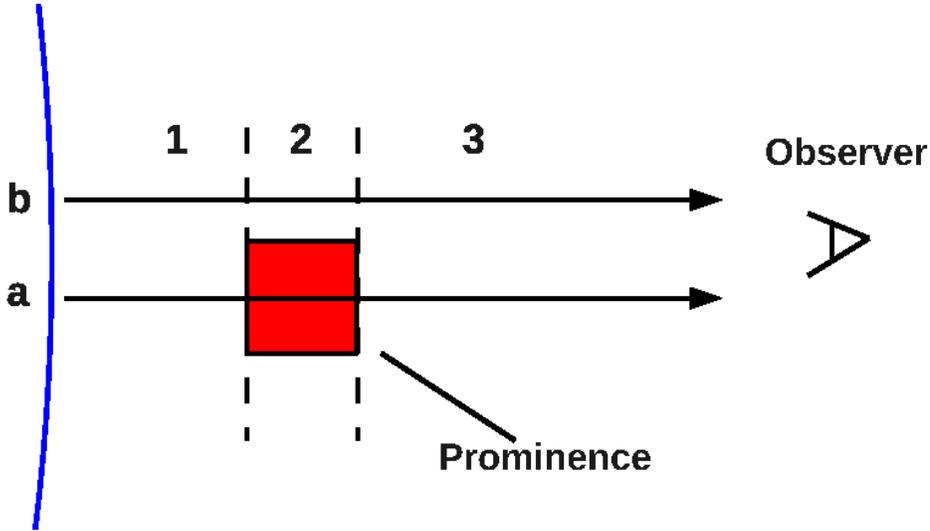}
\caption{\label{geometry} Cartoon showing a simplified version of the
geometry discussed in Section~\ref{applications}. Regions~1, 2 and 3 
define the ``background'', ``prominence'' and ``foreground'' regions, 
respectively, of both lines of sight $a$ and $b$.}
\end{figure}

Figure~\ref{geometry} shows a simplified version of the geometry 
of the problem. It is the same as in Gilbert \etal (2005). Let 
$F_a$ and $F_b$ be the measured EUV intensity values observed 
along the lines of sight $a$ and $b$, where the $a$ intercepts 
a prominence while $b$ lies close, but outside, of it. $F_a$ 
and $F_b$ can be given either by spectral line intensities or 
narrow-band images. Both lines of sight are divided in three 
sections, where section~1 indicates the region below the 
prominence (the ``background region''), section~2 corresponds 
to the finite length $S$ of the prominence, and section~3 
covers the entire distance between the upper boundary of the 
prominence and the observer (the ``foreground region''). The 
fluxes $F_a$ and $F_b$ are given by

\begin{eqnarray}
F_a & = & F_{a1}e^{-\tau} + F_{a2} + F_{a3} \nonumber \\
F_b & = & F_{b1} + F_{b2} + F_{b3} \label{flux1}
\end{eqnarray}

\noindent
Our goal is to determine $\tau$ as a function of the two observables
$F_a$ and $F_b$. However, this equation has six more unknowns, namely 
the background, prominence and foreground intensities for each of the
two line of sights, so that some assumptions are needed. First, we 
note that Gilbert \etal (2005) showed that the emission of the prominence 
itself ($F_{a2}$) at coronal temperatures is negligible. Also, when the 
selected lines of sight $a$ and $b$ lie close to each other, and
the prominence or filament is observed far from complex plasma 
configurations their foreground and background emission are likely 
to be similar: $F_{a1}=F_{b1}=F_1$ and $F_{a3}=F_{b3}=F_3$. Thus, 
the set of Equations~\ref{flux1} simplifies to

\begin{eqnarray}
F_a & = & F_{1}e^{-\tau} + F_{3} \nonumber \\
F_b & = & F_{1} + F_{2} + F_{3} \label{flux2}
\end{eqnarray}

\noindent
where we also indicate $F_2=F_{b2}$ for convenience of notation.

\subsection{Special case I}

If the absorbing material is located at altitudes much larger than 
the scale height of the coronal emission in the solar atmosphere, 
like for example an accelerating erupting prominence during a CME 
onset observed against the solar disk, the emission from the foreground 
and from the plasma in section~2 of the line of sight becomes negligible 
relative to the background emission. This is due to the fact that 
the plasma density at or above the absorbing material height is 
low. In this simplified case, $F_2 \ll F_1$ and $F_3 \ll F_1$, so 
that 

\begin{eqnarray}
\tau & = & -\ln {{F_a}\over{F_b}} \nonumber \\
L{\left({T}\right)} & = & -{{1}\over{k_{eff}{\left({T}\right)}}} \ln{\left({{{F_a}\over{F_b}}}\right)}
\label{case1}
\end{eqnarray}

\noindent
and the diagnostic technique outlined in the previous section can 
be directly applied. When time series of observations are available, 
this technique can provide the measurement of the erupting prominence 
temperature as a function of time as long as the prominence is 
absorbing and, if enough data are available and the plasma 
temperature is in the right range, also the He/H relative abundance.

\subsection{Special case II}

In this case we consider a quiescent prominence sitting at the solar
limb in the absence of active regions in the foreground and background.
Under this configuration, the foreground and background emission can 
be assumed to be roughly the same as their line of sight length and 
plasma physical conditions are approximately similar: $F_1 \simeq F_3$. 
Also, the length of the line of sight at the limb is much larger than 
the prominence depth $S$, so that $F_2$ can be safely assumed to be 
negligible even if it is close to the limb. Simple algebraic 
considerations allow us to rewrite Equations~\ref{flux1} as 

\begin{eqnarray}
\tau & = & -\ln {\left({2{{F_a}\over{F_b}} - 1}\right)} \nonumber \\
L{\left({T}\right)} & = & -{{1}\over{k_{eff}{\left({T}\right)}}} \ln {\left({2{{F_a}\over{F_b}} - 1}\right)}
\label{case2}
\end{eqnarray}

\noindent
and apply the diagnostic technique described in the previous section
to the $L{\left({T}\right)}$ functions.

\subsection{General case}
 
Equations~\ref{flux2} have two observables and four unknown quantities, 
so that some assumptions (or more observables) are needed. To deal with 
this case, we follow the same approach as Gilbert \etal (2005). They 
considered a configuration where two very close parts of the prominence 
were observed against two very different backgrounds, and assumed that 
the absorption term $e^{-\tau}$ is the same in both locations. This 
configuration is very easily obtained when the same prominence is 
observed across the solar limb, so that a portion of the prominence 
can be studied on the disk (we will call this position $D$), and 
another portion lies outside the limb (position $L$).

Gilbert \etal (2005) further assumed that 1) $F_2$ was proportional to 
the foreground intensity $F_3$, so that $F_2 = \gamma F_3$ everywhere 
in proximity of the prominence, and 2) that the foreground emission of 
the $D$ and $L$ regions have a constant ratio $F_3^D/F_3^L = \beta$. 
Under these assumptions, we have four different observables:

\begin{eqnarray}
F_a^L & = & F_1^L + (\gamma+1)F_3^L \nonumber \\
F_b^L & = & F_1^L e^{-\tau} + F_3^L \\
F_a^D & = & F_1^D + \beta (\gamma+1)F_3^L \nonumber \\
F_b^D & = & F_1^D e^{-\tau} + \beta F_3^L \nonumber
\label{main_system}
\end{eqnarray}

\noindent
where $F_a^D, F_b^D, F_a^L$ and $F_b^L$ are observed from spectrally
resolved or narrow band filter images. Simple algebraic consideration 
allow us to estimate the coefficient $e^{-\tau}$ as

\begin{equation}
e^{-\tau} = {{F_b^L - \beta F_b^D}\over{F_a^L - \beta F_a^D}}
\label{gen-case}
\end{equation}

\noindent
The advantage of Equation~\ref{gen-case} is that it requires only the
estimation of the constant $\beta$ and not of $\gamma$. Gilbert \etal
(2005) determined $\beta$ assuming that the path length difference 
between positions $L$ and $D$ was negligible so that the only difference
between the two positions was due to the intensity falloff with distance
from the limb, which can be easily determined from the observations
themselves.

Equation~\ref{gen-case} can be applied to any spectral line or narrow
band image for which two positions $L$ and $D$ can be selected, so that

\begin{eqnarray}
\tau & = & -\ln {\left({{{F_b^L - \beta F_b^D}\over{F_a^L - \beta F_a^D}}}\right)} \nonumber \\
L{\left({T}\right)} & = & -{{1}\over{k_{eff}{\left({T}\right)}}} \ln {\left({{{F_b^L - \beta F_b^D}\over{F_a^L - \beta F_a^D}}}\right)}
\label{case3}
\end{eqnarray}

\noindent
and the diagnostic technique can be applied.

\section{Conclusions}
\label{summary}

The diagnostic technique that we have developed in this work can
be very useful in several occasions. If absorption from prominence
plasmas with temperature lower than $\simeq 15,000$~K is considered,
the absorption coefficient does not depend on the temperature, so
that the availability of lines in different spectral regions can 
provide an estimate of the He abundance relative to H ($A_{He}$) 
even without the knowledge of the thermal properties of the 
prominence itself.

For hotter prominences, different combinations of observations
in the 3 spectral regions where absorption of different species
is dominant can provide very precise temperature estimates, along
with He/H and $N_H$ determinations. These properties are most
important in studies of erupting prominences, where the prominence
plasma is heated to temperatures 10 or more times typical quiescent
values, and knowledge of the temporal evolution of the temperature
can provide vital constraints to models of CME acceleration and
heating. Also, measurements of $A_{He}$ can provide a direct,
quantitative link between remote observations of erupting prominences
near the Sun and in-situ measurements of the properties of prominence 
plasmas in the core of ICMEs.

There are two main advantages in using this method. First, the solar
structures that can be studied with our technique can be easily
identified and studied for long periods of time, especially when 
images from several channels at different wavelengths are available.
Second, the diagnostic technique itself is very fast and easy to
apply to large datasets, so that it allows rapid and accurate
measurements of column density and plasma temperature as a function
of time over extended areas. This makes this technique ideal both to
study individual filaments and CME eventss, as well as to make 
systematic surveys on the physical properties of many events. Also,
if a reasonable assumption or estimate on the geometry of the absorbing
plasma can be made, this method provides a direct and relatively accurate
measurement of the plasma density.

The present diagnostic technique can be applied to data from all the
recent and current space missions, such as SOHO, TRACE, STEREO, Hinode
and SDO, and to future ones such as Solar-C, Solar Orbiter, and Solar
Probe.

\acknowledgements

The work of EL is supported by the NNX11AC20G and NNX10AQ58G NASA
grants, and by NSF grant AGS-1154443. Fabio Reale acknowledges support 
from Italian Ministero dell'Universit\'a e Ricerca and from Agenzia 
Spaziale Italiana (ASI), contract I/015/07/0.

\begin{table}[!t]
\begin{center}
\begin{tabular}{ccll}
Region & Wvl. range & Spectrometer & Imager \\
\hline
 & & & \\
 A & $\lambda > 912$           & SUMER      &                       \\
 B & $504 < \lambda < 912$     & CDS, SUMER &                       \\
 C & $228 < \lambda < 504$     & CDS, EIS   & EIT, EUVI, TRACE, AIA \\
 D & $100 < \lambda < 228$     & EIS        & EIT, EUVI, TRACE, AIA \\
 & & & \\
\hline
\end{tabular}
\end{center}
\caption{\label{combinations} Space-borne instruments observing EUV
radiation in the four spectral regions between 100~\AA\ and 1000~\AA\
characterized by different \ion[H i], \ion[He i] and \ion[He ii]
absorption properties.}
\end{table}

\end{document}